\documentclass[10pt,twocolumn,letterpaper]{article}

\usepackage{iccv}
\usepackage{times}
\usepackage{epsfig}
\usepackage{graphicx}
\usepackage{amsmath}
\usepackage{amssymb}
\usepackage{booktabs}
\usepackage[accsupp]{axessibility}  

\usepackage[breaklinks=true,bookmarks=false]{hyperref}

\newcommand\blfootnote[1]{%
  \begingroup
  \renewcommand\thefootnote{}\footnote{#1}%
  \addtocounter{footnote}{-1}%
  \endgroup
}

\iccvfinalcopy 

\def\iccvPaperID{7} 

\ificcvfinal\fi

\begin{document}

\title{Simultaneous Nuclear Instance and Layer Segmentation in Oral Epithelial Dysplasia}

\author{Adam J. Shephard\textsuperscript{1}, Simon Graham\textsuperscript{1}, R.M. Saad Bashir\textsuperscript{1}, Mostafa Jahanifar\textsuperscript{1},\\
Hanya Mahmood\textsuperscript{2}, Syed Ali Khurram\textsuperscript{2}, Nasir M. Rajpoot\textsuperscript{1}\\
\\
\textsuperscript{1}Department of Computer Science, University of Warwick, Coventry, UK\\
\textsuperscript{2}School of Clinical Dentistry, University of Sheffield, Sheffield, UK \\
{\tt\small adam.shephard@warwick.ac.uk}

}

\maketitle
\ificcvfinal\thispagestyle{empty}\fi

\begin{abstract}
    Oral epithelial dysplasia (OED) is a pre-malignant histopathological diagnosis given to lesions of the oral cavity. Predicting OED grade or whether a case will transition to malignancy is critical for early detection and appropriate treatment. OED typically begins in the lower third of the epithelium before progressing upwards with grade severity, thus we have suggested that segmenting intra-epithelial layers, in addition to individual nuclei, may enable researchers to evaluate important layer-specific morphological features for grade/malignancy prediction. We present HoVer-Net+, a deep learning framework to simultaneously segment (and classify) nuclei and (intra-)epithelial layers in H\&E stained slides from OED cases. The proposed architecture consists of an encoder branch and four decoder branches for simultaneous instance segmentation of nuclei and semantic segmentation of the epithelial layers. We show that the proposed model achieves the state-of-the-art (SOTA) performance in both tasks, with no additional costs when compared to previous SOTA methods for each task. To the best of our knowledge, ours is the first method for simultaneous nuclear instance segmentation and semantic tissue segmentation, with potential for use in computational pathology for other similar simultaneous tasks and for future studies into malignancy prediction.
\end{abstract}

\section{Introduction}

Head and neck cancer is among the top 10 most common cancers worldwide, with an estimated incidence of 150,000 new cases every year in Europe alone \cite{EuropeanCancerPatientCoalitionEuropeanCancers}. These cancers are usually detected late, with 60\% of new cases presenting with an advanced stage at diagnosis. Prognosis for advanced stages is poor, having a five-year survival rate of just 40\%. This increases drastically with early diagnosis to 80-90\% \cite{EuropeanCancerPatientCoalitionEuropeanCancers}. Head and neck cancers encompass a large group of cancers; however, oral squamous cell carcinomas (OSCCs) of the oral cavity are some of the most common. 

Clinically, oral cancers often present as white or red lesions of the oral mucosa. Such lesions are typically preceded by a pre-cancerous state, a group of lesions termed oral potentially malignant disorders (OPMDs) including homogeneous/non-homogeneous leucoplakia (white/white-red speckled lesions) or erythroplakia (red lesions) \cite{Speight2018OralMalignancy}. Biopsies can be taken of these lesions allowing detailed microscopic examination by pathologists to determine the presence or absence of pre-cancer (oral epithelial dysplasia/OED) or cancer. Lesions of the oral mucosa exhibiting dysplasia are statistically more likely to transition into OSCC than non-dysplastic lesions \cite{Liu2011MalignantCases}. Other risk factors of malignancy transition include sex, site/type of lesion, and habits (\eg tobacco and alcohol consumption) \cite{Speight2018OralMalignancy}.

OED lesions can be diagnosed and graded based on a variety of different classification systems, with the criteria set out by the World Health Organisation (WHO) in 2017 \cite{Takata2017TumoursLesions} being considered the gold standard, which grades OED lesions as mild, moderate or severely dysplastic based on cytological and architectural features \cite{Takata2017TumoursLesions}. Although severely dysplastic cases are most likely to become cancerous, progression has also been reported in OED with lower grades (\ie mild and moderate). This could be attributed to the fact that OED grading is subjective with evidence of both inter- and intra-rater variability \cite{Speight2018OralMalignancy,Kujan2007WhyVariation,Shubhasini2017Inter-Dysplasia}.\blfootnote {\textcopyright \space 2021 IEEE. Personal use of this material is permitted. Permission from IEEE must be obtained for all other uses, in any current or future media, including reprinting/republishing this material for advertising or promotional purposes, creating new collective works, for resale or redistribution to servers or lists, or reuse of any copyrighted component of this work in other works.} This lack of reproducibility may lead to over-treatment or in most cases insufficient treatment. This is critical as appropriate treatment and transformation prediction (based on correct grading) can stop a cancer from developing or allow treatment at an early stage conferring better prognosis.

Recent advances in digital pathology have allowed the digitisation of histology slides as whole slide images (WSIs) using high-resolution digital scanners. This has also led to a significant growth in the field of computational pathology \cite{Bera2019ArtificialOncology}. The parallel advances of new deep learning techniques have complimented both pathology and radiology, allowing the automation of pipelines and displaying the promise of deep learning for predicting patient outcomes \cite{Bera2019ArtificialOncology,vanderLaak2021DeepClinic}. In pathology, deep learning has been used to automatically segment the epithelium within a range of histology images (\eg oral, cervical, prostate) \cite{Sornapudi2020EpithNet:Images,Bulten2019EpitheliumStandard, Bashir2020AutomatedImages} and to further segment and classify individual nuclei within WSIs \cite{Graham2019Hover-Net:Images, Naylor2019SegmentationMap}. Within OED, it has been used to segment the epithelium into sub-regions: the lower basal layer, the middle epithelial layer and the superior keratin layer \cite{Bashir2020AutomatedImages}. Initial OED changes typically manifest in the lower third of the epithelium – encompassing the basal layer – progressing upwards into the middle and upper thirds of the epithelium as the grade worsens. Thus, it has been proposed that segmentation of the different epithelial layers may allow the generation of layer-specific features (\eg layer widths) that may aid in OED grading and in predicting malignant progression \cite{Bashir2020AutomatedImages}. Deep learning tools therefore provide a potential avenue for reducing grading variability, whilst ensuring cross-site consistency in informing treatment decisions \cite{Mahmood2020UseReview}.

In this work, we suggest that the segmentation of both the individual nuclei and the three layers of the epithelium are important steps in any downstream analyses of dysplasia in the epithelium, not necessarily confined to oral epithelium. We hypothesise that these tasks are complementary and may aid each individual segmentation/classification task. However, manual segmentation of individual nuclei in WSIs is time-consuming, thus collecting such annotations at a large-scale provides a unique challenge. A range of automatic deep learning techniques have been proposed for this task \cite{Graham2019Hover-Net:Images,Naylor2019SegmentationMap, Raza2019Micro-Net:Images}; however, most methods struggle to differentiate nuclei in crowded scenes or in new images. Semi-automated deep learning techniques \cite{AlemiKoohbanani2020NuClick:Images} have been proposed to combat this problem, by manually providing a seed for each nucleus. Due to the complexity in generating ground truth (GT) nuclear annotations, we propose using a semi-automated multi-stage deep learning framework with manual refinement to generate our GT nuclei annotations. Finally, we present a deep learning framework, \textit{HoVer-Net+}, for simultaneous nuclear instance segmentation (and classification) and semantic segmentation of the layers of the epithelium based on Haematoxylin and Eosin (H\&E) stained histopathology slides of the oral mucosa in OED. We believe this to be the first approach to combine both nuclear and layer segmentation in a single pipeline.

\section{Related Work}

\subsection{Nuclear Instance Segmentation/Classification}

Multiple approaches have been proposed for the segmentation of nuclear instances on WSIs, using traditional machine learning techniques such as watershed-based techniques, active contours and pixel classification \cite{Gao2016HierarchicalImages,Veta2013AutomaticImages, Mouelhi2013AutomaticMethod}. In recent years, deep learning methods have proven  to be most common, due to their popularity and superior performance achieved throughout the computer vision literature for segmentation \cite{Litjens2017AAnalysis,Ronneberger2015U-net:Segmentation, Long2015FullySegmentation}. Deep learning models extract a hierarchy of features that are used to inform the final segmentation. U-Net is a popular approach consisting of an encoder and a decoder with multiple skip connections between layers \cite{Ronneberger2015U-net:Segmentation}. It has been used for semantic segmentation in a plethora of different tasks, and has since been adapted to segment nuclear instances by the incorporation of a weighted loss function. However, a criticism of U-Net is its inability to efficiently disentangle touching or overlapping nuclei. Micro-Net is another deep learning network that has been used for the segmentation of nuclei in histology images \cite{Raza2019Micro-Net:Images}. This model combines multiple image resolutions to achieve superior results.

More recent methods have attempted to improve nuclear instance segmentation performance by treating the task as a regression problem. DIST formulates the problem as a regression task of the distance map \cite{Naylor2019SegmentationMap}, where the distance map gives the distance of a nuclear pixel to the centre of mass of the nuclear instance. In \cite{Graham2019Hover-Net:Images}, the authors incorporate a novel HoVer branch in their encoder-decoder framework that estimates the horizontal and vertical distances of each nuclear pixel from their respective nucleus’ centre of mass. This results in large gradients at the shared boundaries of touching instances, thus allowing more efficient nuclear edge discrimination. HoVer-Net comprises of a single encoder branch and three decoder branches for simultaneous nuclear instance segmentation and classification and has achieved state-of-the-art (SOTA) performance on multiple H\&E stained histology image datasets (\eg Kumar \cite{Kumar2017APathology} and CoNSeP \cite{Graham2019Hover-Net:Images} \etc). This demonstrates its transferability to different histology images/pathologies. In this work, HoVer-Net is employed for nuclear segmentation and classification. We further hypothesise that the incorporation of an additional decoder branch for layer segmentation may aid in improving both nuclear and layer classification.

In \cite{AlemiKoohbanani2020NuClick:Images}, the authors  proposed a supervised framework NuClick that can interactively segment nuclear instances provided only with a single point (click) inside each nucleus. The encoder-decoder design of the NuClick architecture, benefiting from multi-scale convolutional blocks and incorporating a weighted loss function, ensures high-quality segmentation of nuclei at various scales even in the cases of touching instances/crowded scenes. The generalisability of this method, due to the provided guiding signals (clicks), enables the users to collect nuclear instance segmentations for a large set of data rapidly and efficiently \cite{AlemiKoohbanani2020NuClick:Images}; however, we suggest that this may be an impractical way of generating instance segmentations for large cohorts of WSIs. In our work, we have therefore resorted to first generating a reliable dataset of patches in which nuclear instances are annotated with NuClick, and then fine-tuning HoVer-Net on that dataset to make it ready for inference at the WSI level.

\begin{figure*}
\begin{center}
\includegraphics[width=16cm]{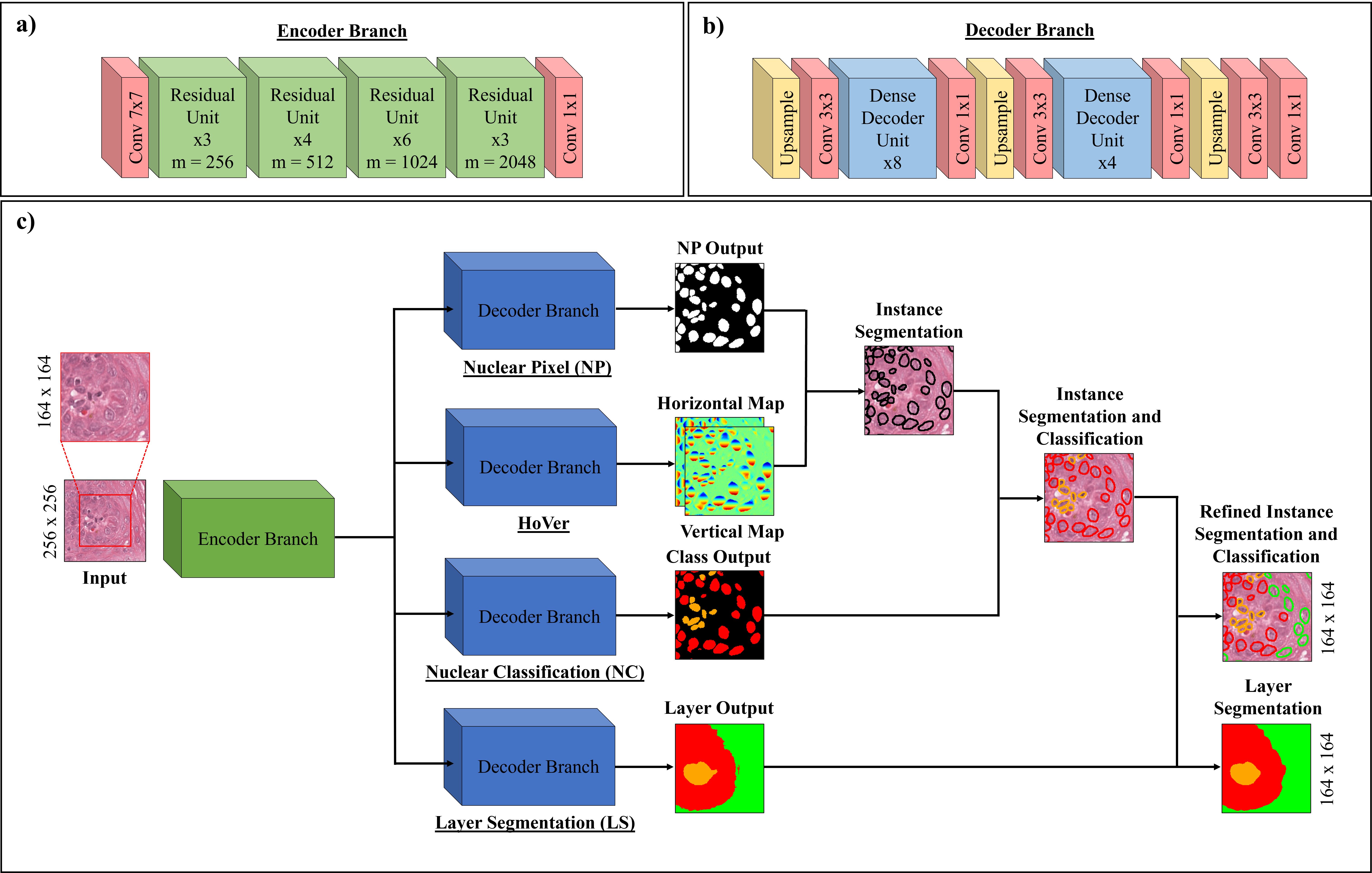}
\end{center}
   \caption{The architecture of HoVer-Net+, including the post-processing steps taken to achieve the final output, showing a) the encoder branch, b) the decoder branch and c) the overall architecture, where m denotes the number of feature maps within the residual units.}
\label{fig:arch}
\end{figure*}

\subsection{Layer Segmentation}

Deep learning approaches have been used extensively for semantic segmentation, with models such as the fully convolutional neural network \cite{Long2015FullySegmentation}, U-Net \cite{Ronneberger2015U-net:Segmentation} and DeepLabv3 \cite{Chen2017RethinkingSegmentation} being considered the SOTA. Multiple works have used these methods to automatically segment the epithelium in histology images. In \cite{Bulten2019EpitheliumStandard}, the authors used U-Net to successfully segment the epithelium in IHC and H\&E stained images. In \cite{Sornapudi2020EpithNet:Images}, the authors presented EpithNet, a variant of VGG \cite{Simonyan2015VeryRecognition}, to segment epithelium in cervical histology images. This method incorporated a novel step approximating edge contours with Bezier curves to outperform U-Net. In \cite{Bashir2020AutomatedImages}, DeepLabv3 was used to further segment oral epithelium into its constituent layers: basal layer, (middle) epithelial layer, (superficial) keratin layer, in order to generate layer-specific morphological features for predicting OED grade.

\section{Methodology}

\subsection{Network Architecture}

The network proposed in this study is based on HoVer-Net \cite{Graham2019Hover-Net:Images}, which performs nuclear instance segmentation and classification. HoVer-Net is a deep learning framework that consists of an encoder branch, and three decoder branches. The encoder branch is based on the pre-activated residual network with 50 layers \cite{He2016IdentityNetworks}, consisting of four consecutive pre-activated residual units. The decoder branches of the network perform nearest-neighbour up-sampling, through up-sampling operations and densely connected units \cite{Huang2017DenselyNetworks}, in conjunction with skip connections from the encoder \cite{Ronneberger2015U-net:Segmentation}. 

HoVer-Net consists of three decoder branches: the nuclear pixel (NP) branch, the horizontal and vertical distances (HoVer) branch, and the nuclear classification (NC) branch. The NP branch performs binary classification, thus predicting whether a pixel is foreground (nucleus) or background. The HoVer branch estimates the horizontal and vertical distances of each pixel in a nucleus from the respective nucleus’ centre of mass, resulting in large gradients at shared nuclear boundaries. Finally, the NC branch predicts the type of nucleus for each pixel. Together, the NP and HoVer branches perform nuclear instance segmentation, whilst the NC branch performs nuclear classification. In this work, we have adapted HoVer-Net to include a fourth decoder branch that acts as the layer segmentation (LS) branch, in an adaption we have named HoVer-Net+. Within this work, we consider a larger input size (256$\times$256) than was used in the original HoVer-Net to speed up inference. Figure \ref{fig:arch} shows an overview of the model architecture.

\subsection{Loss Function}

The same overall loss function as utilised in \cite{Graham2019Hover-Net:Images} was used, with the addition of two extra terms relating to the LS branch. The proposed method consists of four decoder branches, each containing weights that are jointly optimised with the following overall loss $\mathfrak{L}$:

\begin{equation}\label{my_first_eqn}
\begin{aligned}
	\mathfrak{L} = & \mathfrak{L}_a\lambda_a + \mathfrak{L}_b\lambda_b + \mathfrak{L}_c \lambda_c + \mathfrak{L}_d\lambda_d \\
	& + \mathfrak{L}_e\lambda_e + \mathfrak{L}_f\lambda_f + \mathfrak{L}_g\lambda_g + \mathfrak{L}_h\lambda_h,
\end{aligned}
\end{equation}

\noindent where $\mathfrak{L}_a$ and $\mathfrak{L}_b$ are regression losses with respect to the output of the HoVer branch,  $\mathfrak{L}_c$ and $\mathfrak{L}_d$ are losses with respect to the output at the NP branch, $\mathfrak{L}_e$ and $\mathfrak{L}_f$ are losses with respect to the output at the NC branch, and $\mathfrak{L}_g$ and $\mathfrak{L}_h$ are losses with respect to the output of the LS branch, and $\lambda_a$ to $\lambda_h$ are weights assigned to each loss function. We assigned $\lambda_b$ a value of 2 in line with \cite{Graham2019Hover-Net:Images}, whilst we found values of 1 best for the remaining weights. 

At the output of the HoVer branch we calculated the mean squared error between the predicted horizontal and vertical distances and the GT ($\mathfrak{L}_a$). We also calculated the mean squared error between the horizontal and vertical gradients of the horizontal and vertical maps respectively, and the corresponding gradients of the GT ($\mathfrak{L}_b$). At the output of the NP, NC and LS branches, we calculated the cross-entropy loss ($\mathfrak{L}_c$, $\mathfrak{L}_e$, $\mathfrak{L}_g$) and the dice loss ($\mathfrak{L}_d$, $\mathfrak{L}_f$, $\mathfrak{L}_h$). Consequently, each branch consists of two loss terms summed together. For the full formulation of the loss terms used in this work, see \cite{Graham2019Hover-Net:Images}.

\subsection{Post-Processing}

The three original branches of HoVer-Net were used to segment and classify nuclei as either epithelial or other (\eg connective/immune cells) nuclei. The same post-processing steps for nuclear instance segmentation and classification were used as in \cite{Graham2019Hover-Net:Images}. This generated instance segmentations using the HoVer and NP branches, combined with a marker-based watershed. Majority voting was used to determine the class of each nucleus via the NC branch. The output layer segmentation of the LS branch was post-processed via the morphological opening and closing of layers, along with the removal of small objects and holes. Any nuclei that were classified as epithelial via the original HoVer-Net architecture were then further differentiated by the associated predicted layer that their centroid was located in. Thus, if a nucleus was predicted to be an epithelial nucleus, then that nucleus would be further sub-classified as a basal, epithelial or keratin nucleus depending on the associated predicted layer of the LS branch. This additionally allows for other nucleus types, such as lymphocytes that are known to penetrate the epithelial layer in OED, to be found in the epithelial layer. Note, the LS branch was used to assign layers to each epithelial nuclei over the NC branch, to ensure a smooth boundary between nuclei of different layers. HoVer-Net+ produces two sets of outputs: nuclear classification maps and layer segmentation maps. Each of these maps consists of 5 classes: background, other, basal, epithelium, keratin. Figure \ref{fig:arch} displays example outputs showing other tissue as orange, basal as red, and epithelium as green. No background or keratin classes are present in the example images. 

\subsection{Evaluation Metrics}

For the evaluation of nuclear instance segmentation and classification we have used the same metrics as used in \cite{Graham2019Hover-Net:Images}. Therefore, nuclear instance segmentation performance was assessed based on Panoptic Quality (PQ), a score originally proposed by \cite{Kirillov2019PanopticSegmentation}, unifying the results of instance segmentation and semantic segmentation. Although PQ provides a good overall metric for comparing models, it is beneficial to interpret this in conjunction with its constituent elements: detection quality (DQ; or F1-score) and segmentation quality (SQ). We further report a Dice score comparing all segmented nuclei against the background, and an aggregated Jaccard index (AJI) \cite{Kumar2017APathology}. PQ and AJI compare each pair of predicted and GT nuclei, classifying a matched pair (true positive) as when their intersection over union is greater than 0.5. For the evaluation of nuclear classification performance, we report the average value over all images for: the F1-score for detection F\textsubscript{d} (\ie all nuclear types); and F1-score for classification F\textsubscript{c} for each of the individual nuclei types (\eg other: F\textsubscript{c}\textsuperscript{o}, basal: F\textsubscript{c}\textsuperscript{b}, epithelium: F\textsubscript{c}\textsuperscript{e}). We refer the reader to \cite{Graham2019Hover-Net:Images} for an in-depth description of these metrics. Finally, for layer segmentation, we report aggregated precision, recall, accuracy and F1-score. We further provide a breakdown of the F1-score by each class studied (\eg F1-score for basal {\em vs.} rest).

\section{Preliminary Experiment}

\subsection{Data}

The study data consisted of 43 WSIs from H\&E stained sections of the oral mucosa attained at the University of Sheffield, UK, and scanned at 20$\times$ (0.4952 microns-per-pixel) using an Aperio CS2 scanner. Of these cases, 38 had OED and 5 were healthy controls. An expert oral and maxillofacial pathologist (SAK) graded the OED cases as either mild, moderate, or severe according to the WHO 2017 classification system \cite{Takata2017TumoursLesions}. Of the 38 cases, 14 were rated as mild, 13 as moderate, and 11 as severe. Analysis of the data was conducted under ethical approval by the NHS Health Research Authority West Midlands titled ``Artificial Intelligence to grade oral epithelial dysplasia and predict malignant transformation'' (Ref: 18/WM/0335). The pathologist further annotated the epithelium into three layers (basal layer, epithelium, superficial (keratin) layer) using the Automated Slide Analysis Platform software (https://computationalpathologygroup.github.io/ASAP/). Data were split into training/validation/testing sets with a ratio of 29/7/7, stratified by OED grade (including normal cases). Tissue masks were generated for each WSI via Otsu thresholding and the further removal of small holes and objects. Layer masks were generated for each WSI by combining the layer segmentations with the tissue mask.

\subsection{Nuclear Ground Truth Generation}

\subsubsection{NuClick}

The manual segmentation of individual nuclei in WSIs is time-consuming and subject to inter/intra-rater variability, thus nuclear instance maps were instead generated for a subset of cases (n = 20) using NuClick \cite{AlemiKoohbanani2020NuClick:Images}. NuClick takes nuclear instance point annotations as inputs and outputs nuclear instance segmentations that are superior to those produced by fully automated approaches (specifically with touching nuclei) \cite{AlemiKoohbanani2020NuClick:Images}. Thus, NuClick is suitable for generating GT annotations. The expert pathologist drew large regions of interest (ROIs) for these 20 cases and assigned a single point to each nucleus. Each instance was assigned a class as either epithelial or other nuclei. Following processing, all NuClick annotations were visually inspected and edited using the GIMP software (https://www.gimp.org/) where necessary. This was generally only necessary in the granular layer of the epithelium. This resulted in 1,139 patches (256$\times$256 at 20$\times$) containing nearly 50,000 nuclei. 

\subsubsection{HoVer-Net}

The refined NuClick annotations were further used to fine-tune the original HoVer-Net (pre-trained on the PanNuke dataset \cite{Gamper2020PanNukeBaselines}) to segment and classify nuclei in OED H\&E images as either epithelial or other nuclei. This model was trained/validated on the manually-refined NuClick patches from the training and validation sets alone (12 and 4 ROIs respectively), to ensure an unbiased result on the test data in the proceeding experiments. The NuClick test set (4 ROIs from 4 cases) was held back for final evaluation in the main experiments. HoVer-Net was implemented with the PyTorch library on a workstation equipped with two NVIDIA RTX 5000 GPUs. It was trained in two phases, with only the decoder branches trained for the first 50 epochs, and with all branches in the next 50 epochs. A batch size of 8 and 4 on each GPU were used across these two phases respectively. The Adam optimizer was used with a learning rate that decayed from 10\textsuperscript{-4} to 10\textsuperscript{-5} after 25 epochs in each phase. The following quality metrics were attained for validation on the patches obtained from the NuClick ROIs for instance segmentation: Dice = 0.715, AJI = 0.669, DQ = 0.798, SQ = 0.710, PQ = 0.562.  For classification we obtained: F\textsubscript{d} = 0.838, F\textsubscript{c}\textsuperscript{o} = 0.821, F\textsubscript{c}\textsuperscript{e} = 0.781. These results demonstrate the success of our NuClick/HoVer-Net pipeline for gaining OED specific GT segmentations.

Following fine-tuning, the trained HoVer-Net model was used for inference on all 43 WSIs in order to produce nuclear instance segmentations for each nucleus in every WSI. We further implemented a quality control step here, checking and manually refining each nuclear class where necessary, to ensure that the classes assigned to each nucleus in all 43 WSIs were correct. The nuclear masks were superimposed on the layer masks for each WSI, to further differentiate epithelial nuclei by intra-epithelial layer (basal, epithelial, keratin). This resulted in nuclear classes corresponding to basal, epithelial, keratin and other nuclei. The other nuclei mainly included nuclei of connective tissue cells, endothelial cells and immune cells. All WSIs were then divided into non-overlapping 256$\times$256 images (252$\times$252 and 268$\times$268 patches for comparative experiments). A balanced dataset was ensured by following the below criteria on patching: a) patch contains some epithelial tissue (any of the three layers); b) patch is greater than 10\% tissue (\ie mean intensity of binary tissue map is $\geq$ 0.1)

\section{Experiments and Results}

\subsection{Data}

\begin{table*}
\begin{center}
\begin{tabular}{lccccccccc}
\hline
Model & Dice & AJI & DQ & SQ  & PQ & F\textsubscript{d} & F\textsubscript{c}\textsuperscript{o} & F\textsubscript{c}\textsuperscript{b}  & F\textsubscript{c}\textsuperscript{e} \\
\hline
DIST & 0.788 &	0.666 &	0.745 &	0.735 &	0.564 & 0.868 &	0.747 & 0.583 &	0.593 \\
U-Net & 0.752 &	0.627 &	0.746 &	0.679 &	0.519 & 0.881 &	0.761 &	0.599	& \textbf{0.598} \\
Micro-Net & \textbf{0.855} &	\textbf{0.787} &	\textbf{0.838} &	0.841 &	0.715 & \textbf{0.892} &	0.777 &	\textbf{0.627} &	0.582 \\
HoVer-Net & 0.848 & 0.775 &	0.823 &	0.866 &	0.725 & 0.890 &	0.761 &	0.600 &	0.580 \\
HoVer-Net+ & 0.849	& 0.778 &	0.825	& \textbf{0.867}	& \textbf{0.727} & 0.890 &	\textbf{0.783} &	0.590 &	0.597 \\
\hline
\end{tabular}
\end{center}
\caption{Comparative experiments for nuclear instance segmentation and classification on the OED dataset.}
\label{table:segm1}
\end{table*}

The preliminary experiment provided GT nuclear and layer segmentations (and classes) for the 43 WSIs used in this study. This resulted in 17,086 patches for training, 3,358 for validation and 4,188 for testing. The efficacy of the proposed HoVer-Net+ model at simultaneous nuclear segmentation (and classification) and layer segmentation, was tested using the data described in the preliminary experiment in a series of further experiments.

\subsection{Implementation}

HoVer-Net+ was tested and compared to other models for nuclear instance segmentation and classification, and layer segmentation. On testing layer segmentation, HoVer-Net+  was further adapted for comparison, to perform layer segmentation only. This trial is labelled HoVer-Net+ (LS). This architecture used a single decoder branch (LS), and aimed to establish whether simultaneous nuclear and layer segmentation improves performance by the incorporation of other histological information. All HoVer-Net(+) models were trained in two phases, with only the decoder branches trained for the first 20 epochs, and with all branches in the next 20 epochs. A batch size of 8 and 4 on each GPU were used across these two phases respectively. The Adam optimizer was used with a learning rate that decayed initially from 10\textsuperscript{-4} to 10\textsuperscript{-5} after 10 epochs in each phase. Model hyper-parameters were found through optimisation on the training/validation set. Further hold-out testing took place after retraining the models on the combined training and validation set. Comparative experiments were performed to test HoVer-Net+'s performance against other benchmark networks for nuclear classification and semantic layer segmentation. For nuclear classification, HoVer-Net+ was tested against U-Net \cite{Ronneberger2015U-net:Segmentation}, DIST \cite{Naylor2019SegmentationMap} and Micro-Net \cite{Raza2019Micro-Net:Images}, where models were trained/tested based on their default parameters, however, where necessary they were adapted to perform classification as described in \cite{Graham2019Hover-Net:Images}. HoVer-Net+ was compared against DeepLabv3 \cite{Chen2017RethinkingSegmentation} and U-Net \cite{Ronneberger2015U-net:Segmentation} for the segmentation of layers, owing to the successes of these models in achieving SOTA results for semantic segmentation. During the training of all models, patches were augmented using flip, rotation, Gaussian blur and median blur. All models were confirmed for convergence after 40 epochs. 

\begin{table}
\begin{center}
\begin{tabular}{lccccc}
\hline
Model & Dice & AJI & DQ & SQ  & PQ \\
\hline
DIST & 0.801 &	0.647 &	0.775 &	0.735 &	0.570\\
U-Net & 0.769 &	0.612 &	0.755 &	0.693 &	0.525 \\
Micro-Net & 0.829 &	0.684 &	0.840 &	0.793 &	0.666 \\
HoVer-Net & 0.836 &	0.691 &	0.860 &	0.801 &	0.689 \\
HoVer-Net+ & \textbf{0.839} &	\textbf{0.694} &	\textbf{0.861} &	\textbf{0.801} &	\textbf{0.690} \\
\hline
\end{tabular}
\end{center}
\caption{Comparative experiments for nuclear instance segmentation on the hold-out manually-refined NuClick OED test set.}
\label{table:segm2}
\end{table}

\begin{figure*}
\begin{center}
\includegraphics[width=17.4cm]{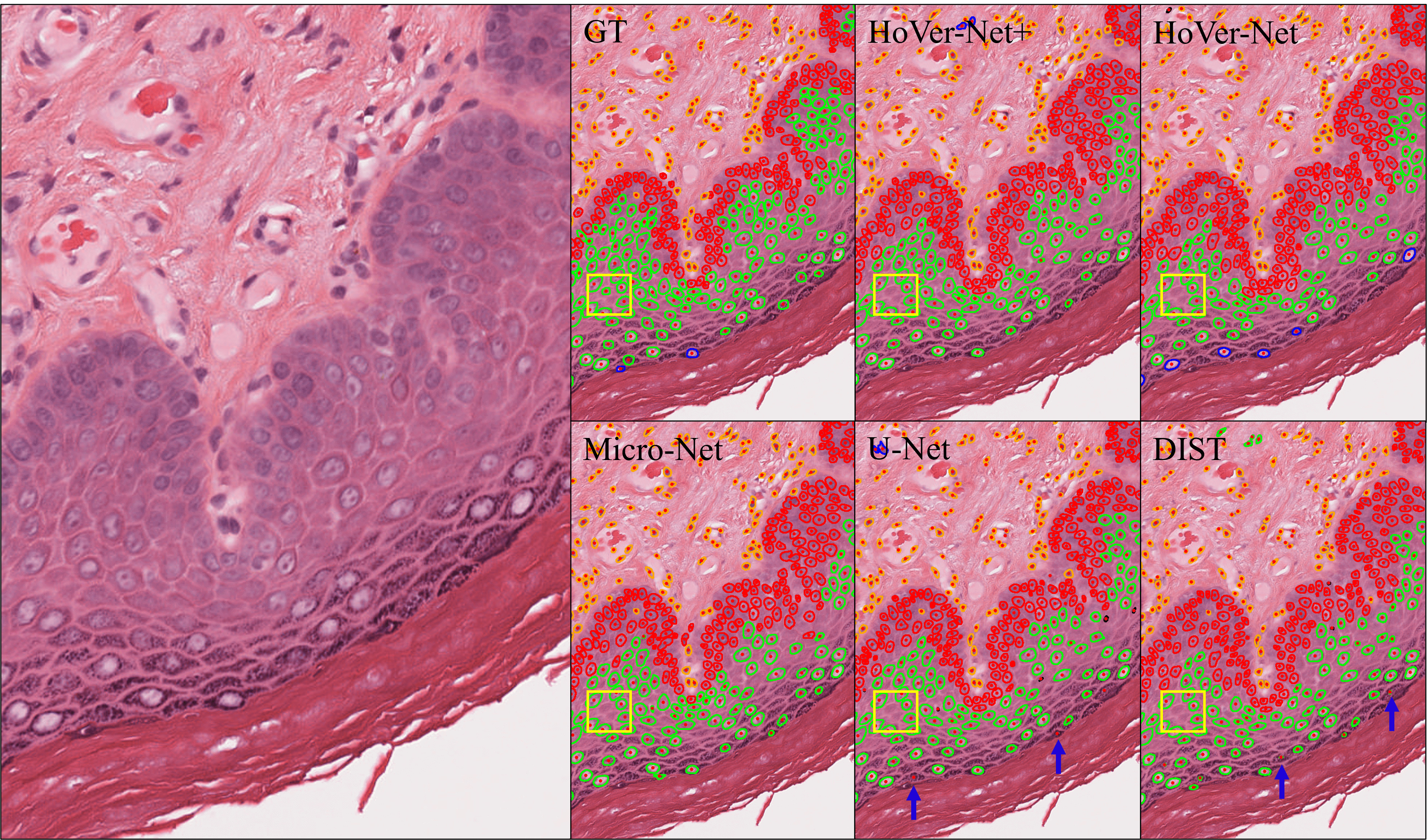}
\end{center}
   \caption{The nuclear segmentations from each of the models compared to the GT. Here, red represents basal, green represents epithelial, blue keratin, and orange other nuclei. The yellow boxes show areas of false negatives when compared to the GT, whilst the blue arrows show false positives.}
\label{fig:nuclei}
\end{figure*}

\subsection{Nuclear Instance Segmentation}

We evaluated HoVer-Net+ for nuclear instance segmentation on the entire OED cohort testing set, by comparing it to the other SOTA approaches. The results for instance segmentation are included in Table \ref{table:segm1}; however, these results will be biased towards the HoVer-Net architectures as HoVer-Net was used to generate the GT instance segmentations, with only the classes manually-refined. Therefore, we further display the results of all models on the hold-out manually-refined NuClick testing set, consisting of four large ROIs from four subjects (see Table \ref{table:segm2}). This test set was not evaluated in the preliminary experiment. Both HoVer-Net and HoVer-Net+ achieved the best performance across both test sets, particularly in relation to PQ. However, Micro-Net was only marginally worse in both scenarios, dissimilar to U-Net and DIST that have a PQ between 0.16 and 0.21 lower than the HoVer-Net models.

\subsection{Nuclear Classification}

Our proposed model classified nuclei as background, other, basal epithelium, epithelium or keratin layer epithelium. Since  few nuclei are generally found in the keratin layer, a metric has not been included for this class. The HoVer-Net(+) models and Micro-Net, have again achieved the highest performance for the segmentation of all nuclei (see Table \ref{table:segm1}), with only minimal differences. However, for classification, HoVer-Net+ has attained the best performance for other nuclei, whilst Micro-Net was best for basal epithelial nuclei. U-Net achieved the highest score for the classification of epithelial nuclei, having a score marginally better than HoVer-Net+. Overall, there was little difference in the performances of the HoVer-Net(+) architectures and Micro-Net for the classification of nuclei, but all performed better than U-Net and DIST. Therefore, simultaneous nuclear and layer segmentation has resulted in a comparative performance with other SOTA approaches, demonstrating the efficiency of our method.

Examination of the quality of each model's segmentation performance compared to the GT in Figure \ref{fig:nuclei} shows how most models have effectively classified each type of nuclei, including other nuclei infiltrating the epithelium. However, it can be seen that U-Net and DIST have regularly segmented small areas of false positives in cells in the granular layer of the epithelium, where no nuclei are visible (see blue arrows). The HoVer-Net(+) and Micro-Net architectures have avoided this. All models have areas of false negatives for some faint nuclei in the epithelium, highlighted by the yellow boxes in Figure \ref{fig:nuclei}.

\subsection{Layer Segmentation}

We further compared HoVer-Net+ to other SOTA semantic segmentation methods: DeepLabv3 and U-Net, for the segmentation of the epithelium in OED (Table \ref{table:layer1}). HoVer-Net+ achieved superior performance for layer segmentation when performing simultaneous layer and nuclear segmentation (LS \& NC) and when performing layer segmentation only (LS) than when compared to both other models. This difference in F1-score was particularly apparent in segmenting the keratin and other tissue; however, was somewhat counteracted in the basal layer, where DeepLabv3 performed the best. Overall, these results, show the superiority of the HoVer-Net+ architecture for semantic segmentation of the epithelium when compared to other SOTA approaches. The marginal difference between HoVer-Net+ for simultaneous layer and nuclear segmentation {\em vs.} for layer segmentation alone, suggest that simultaneous segmentation provides a more efficient approach to these tasks.

A comparison of the layer segmentations of HoVer-Net+ (NC \& LS) {\em vs.} the GT on a ROI from the test set can be seen in Figure 3. The red boxes show areas where HoVer-Net+ has segmented regions that the GT segmentation has misclassified, whilst the blue boxes show regions where HoVer-Net+ has misclassified some basal layer epithelium and other tissue. There are also areas of connective tissue that have been misclassified as keratin. Overall, the predicted segmentations appear to be smoother than the GT.

\begin{table*}
\begin{center}
\begin{tabular}{lccccccccc}
\hline
& \multicolumn{6}{c}{F1-score} &&& \\\cmidrule{2-7}%
Model & Bkgd. & Other & Basal & Epith. & Keratin & Mean & Prec. & Recall & Acc. \\
\hline
U-Net & 0.874 &	0.666	& 0.719 &	0.740	& 0.494 &	0.699 &	0.698 &	0.720 &	0.721\\
DeepLabv3 & 0.863 &	0.751 &	\textbf{0.748} &	0.832 &	0.601 &	0.759 &	0.755 &	0.768 &	0.786\\
HoVer-Net+\textsubscript{LS} & \textbf{0.900} &	\textbf{0.876} &	0.718 &	\textbf{0.868} &	\textbf{0.760} &	\textbf{0.824} & \textbf{0.834} &	0.819 &	\textbf{0.841}\\
HoVer-Net+\textsubscript{LS,NC} & 0.898	& 0.871 &	0.720 &	0.860 &	0.752 &	0.820 &	0.824 &	\textbf{0.821} &	0.835  \\
\hline
\end{tabular}
\end{center}
\caption{Comparative experiments for layer segmentation on the OED dataset.}
\label{table:layer1}
\end{table*}

\begin{figure*}
\begin{center}
\includegraphics[width=17cm]{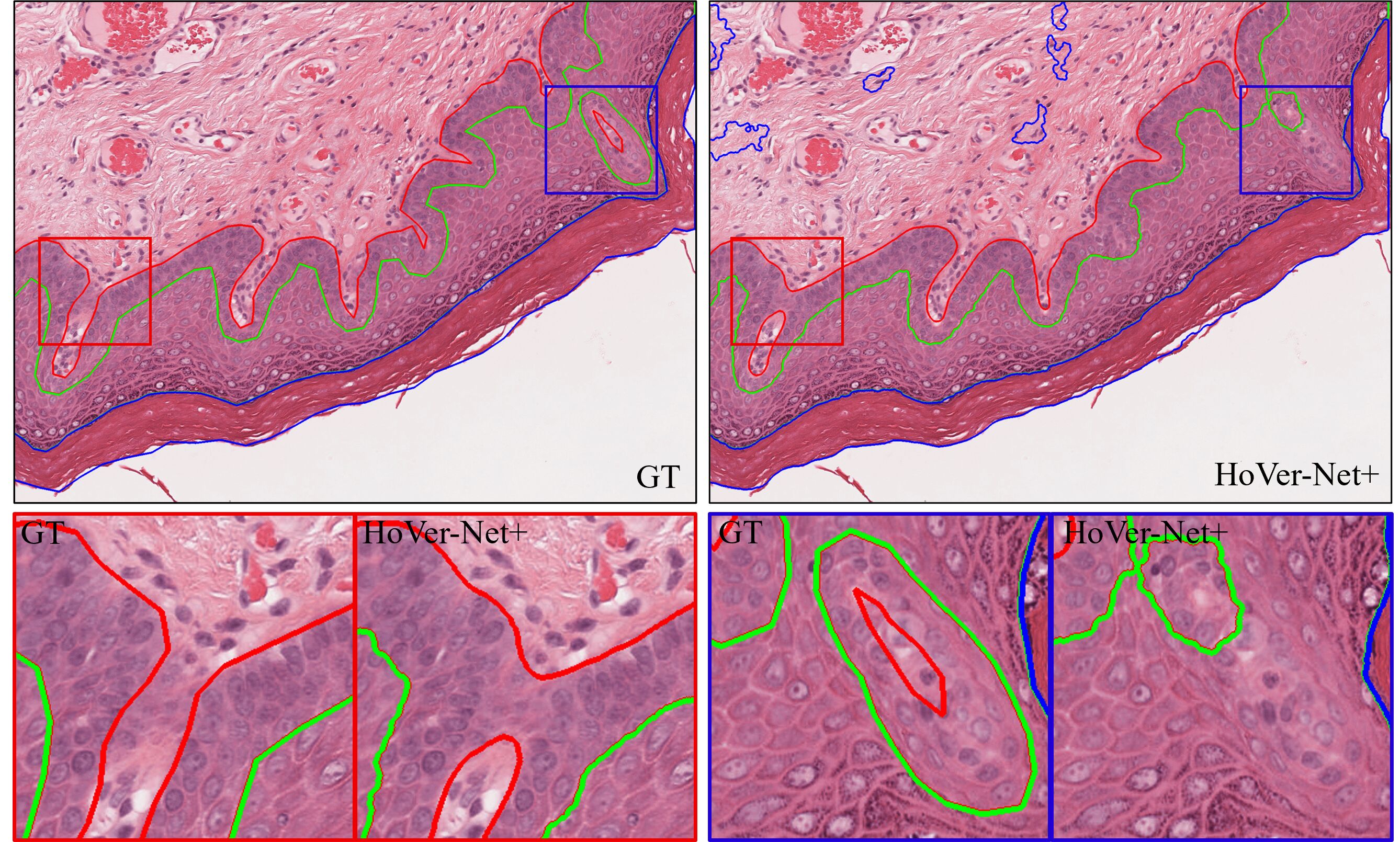}
\end{center}
   \caption{A comparison of the GT {\em vs.} the HoVer-Net+ layer segmentation. In the images, the red line represents the basal layer, the green epithelium, and blue keratin. For clarity, other tissue regions have been excluded from these results.}
\label{fig:layers}
\end{figure*}

\section{Discussion and Conclusion}

In this work we have proposed HoVer-Net+, a new model for simultaneous nuclear instance segmentation (and classification) and semantic segmentation of three layers within the epithelium of OED WSIs. The accurate segmentation of both nuclei and layers within OED cases is a vital step in down-stream analysis, due to the importance of finding layer-specific morphological features for the prediction of  grade and outcome/malignancy. Indeed, previous work by \cite{Bashir2020AutomatedImages} used such features to predict OED grade; however, was limited by a small cohort. The proposed method provides an avenue to perform rapid segmentation of nuclei/layers, allowing studies to scale-up approaches for outcome prediction with large cohorts. However, we emphasise that the potential of the proposed method is not confined to oral epithelial layers, or indeed the epithelium. 

This study has shown HoVer-Net+ to perform in line with the current SOTA approaches for both nuclear segmentation/classification and layer segmentation. For layer segmentation, we found HoVer-Net+ to achieve better results than both U-Net and DeepLabv3. We suggest that this may be due to our experiments taking place at 20$\times$, a necessity for ensuring accurate nuclei segmentation, as opposed to 10$\times$, which has been used in previous work \cite{Sornapudi2020EpithNet:Images,Bashir2020AutomatedImages}. We have shown that simultaneous segmentation of both nuclei and layers can yield SOTA performance by combining both tasks without any loss in performance. This further suggests that important domain-specific features are learnt by the encoder that are useful for both nuclear and layer segmentation. It also demonstrates the adaptability and efficiency of the proposed network for multi-task learning.

This study was limited in its ability to assess nuclear instance segmentation based on the whole cohort, due to the GT nuclear instances being produced by a combination of NuClick and HoVer-Net. However, this method was chosen since the manual segmentation of each nucleus in all 43 WSIs would have been too time consuming. We did, however, provide an unbiased assessment of nuclear segmentation performance by testing each model on the hold-out NuClick ROIs, where HoVer-Net+ still obtained the best performance. Future work should further separate nuclei into more classes (\eg immune and connective cells).

We believe this work advances the field of computational pathology by presenting an efficient model for the simultaneous segmentation of nuclear instances and epithelial layers. We have demonstrated its success in a complex cohort of OED cases, and emphasize its potential in other cohorts for a multitude of tasks. HoVer-Net+ could be used as an essential step to scale-up future studies for the rapid generation of morphological features for use in down-stream analyses to better understand the tumour microenvironment and ultimately improve outcome prediction. Future work could instead use the LS branch to simultaneously segment nuclei and tumour/stroma regions, or other similar simultaneous tasks, more efficiently by incorporating knowledge of underlying cytological changes between the different tissue constituents as captured by the proposed model.

\section*{Acknowledgements}
This work was supported by a Cancer Research UK Early Detection Project Grant, as part of the ANTICIPATE study.

{\small
\bibliographystyle{ieee_fullname}
\bibliography{references.bib}
}

\end{document}